\documentclass{llncs}
\usepackage{makeidx}  
\usepackage[pdftex]{graphicx}
\input epsf

\begin{document}
%
%
\pagestyle{headings}  
%
%

\title{Uniform Linked Lists Contraction}
\titlerunning{Linked List}  
%
\author{Yijie Han}
\authorrunning{Han} 
%
%
\institute{School of Computing and Engineering\\ 
University of Missouri at Kansas City\\
Kansas City, MO 64110, USA\\
\email{hanyij@umkc.edu},\\
}

\maketitle  

\begin{abstract}
We present a parallel algorithm (EREW PRAM algorithm) for linked lists contraction. We show that when
we contract a linked list from size $n$ to size $n/c$ for a suitable constant
$c$ we can pack the linked list into an array of size $n/d$ for a constant $1 < d\leq c$ in the time of 3 coloring the list. Thus for a set of 
linked lists with a total of $n$ elements and the longest list 
has $l$ elements our algorithm contracts them in 
$O(n\log i/p+(\log^{(i)}n+\log i )\log \log l+ \log l)$ time, for an arbitrary 
constructible integer $i$, with $p$ processors on the EREW PRAM, 
where $\log^{(1)} n =\log n$ and $\log^{(t)}n=\log \log^{(t-1)} n$ and $\log^*n=\min \{ i|\log^{(i)} n < 10\}$. 
When $i$ is a constant we get time $O(n/p+\log^{(i)}n\log \log l+\log l)$.
Thus when $l=\Omega (\log^{(c)}n)$ for any constant $c$ we achieve $O(n/p+\log l)$ time. 
The previous best deterministic EREW PRAM algorithm has time $O(n/p+\log n)$ and best CRCW PRAM algorithm has time $O(n/p+\log n/\log \log n+\log l)$.
\end{abstract}

\baselineskip=5.3mm

\section{Introduction}

Linked list contraction refers to the pairing of neighboring nodes on 
the linked list and replacing the two nodes with a newly created node 
in order to have the linked list contracted. In parallel computation we 
need to do this pairing-off for many pairs of nodes concurrently such 
that in one round the size of the linked list shrinks from 
$n$ to $n/c$ for a constant $c>1$. Linked list contraction requires 
this process repeated such that the original linked list is finally 
contracted to a single node (actually because of parallel algorithm design
we need the initial input $n$ nodes linked list to be contracted to
$n/\log n$ nodes as pointer jumping \cite{Wyllie} can be applied after
the linked list has been contracted to $n/\log n$ nodes). Linked list contraction can be used to 
rank the nodes in the linked list such that nodes along the linked 
list receiving numbers $1, 2, 3, . . .,n$. Linked list contraction 
and/or ranking can then be used for many other computations for 
linked lists, trees, and graphs. Thus linked list contraction is 
regarded as a basic or fundamental computation in parallel computation.

The model we use is the EREW (Exclusive Read Exclusive Write) PRAM (Parallel
Random Access Machine) \cite{JaJa,KR}. PRAM is a machine where memory is shared
among all processors. In one step on the EREW (CREW (Concurrent Read Exclusive
Write), CRCW (Concurrent Read Concurrent Write)) PRAM multiple
processors cannot (can, can) read the same memory cell and cannot (cannot,
can) write the same memory cell. When concurrent write happens on
the CRCW PRAM an arbitration rule needs to be used to determine what content
is written into the memory cell. On the ARBITRARY CRCW PRAM an arbitrary processor will succeed in writing when concurrent write happens. On the COMMON
CRCW PRAM processors write into the same memory cell in a step have to
write the same value and this value is written into the memory cell. 

Early work on linked list contraction include the work of Wyllie \cite{Wyllie}
and Kruskal et al. \cite{Kruskal}.
Han achieved $O(n/p+\log n)$ time for linked list contraction of $n$ nodes
on the CRCW PRAM with $p$ processors \cite{HanThesis}.
At the time it was difficult to achieve $O(n/p+\log n)$ time on the EREW PRAM
because it was not known how to do uniform linked list contraction and dynamically balancing the load is difficult.
In 1991 Anderson and Miller came up with a very clever accounting scheme
to calculate the dynamic load and 
obtained an EREW PRAM algorithm \cite{AM} for 
linked list ranking (contraction) with time $O(n/p+\log n)$ 
(conference version of their paper was published in 1988). The existence of the $\log n$ additive factor in 
the complexity of their algorithm comes from two causes. One is 
the pointer jumping \cite{Wyllie} and the other reason is the dynamic load balancing 
scheme of their algorithm because their algorithm requires the contracted linked list be packed before the final pointer jumping stage \cite{Wyllie}. We are unable to reduce the time complexity 
for pointer jumping and therefore in the case of contracting a single 
linked list of size $n$ we cannot improve on the $O(n/p+\log n)$ time 
complexity of Anderson and Miller's algorithm. However, if we are 
contracting many linked lists stored in an array of size $n$ and the 
longest linked list has size $l$ then the only application of pointer 
jumping \cite{Wyllie} will rank the linked lists in $O(n\log l/p+\log l)$ time. 
Therefore, the open question is that in the case of contracting or 
ranking multiple linked lists can we do better. The ideal 
case here is to achieve 
$O(n/p+\log l)$ time. Anderson and Miller's algorithm cannot 
achieve this because their load balancing scheme cannot guarantee 
uniform contraction. Here we use uniform contraction to mean when 
linked lists are contracted they can be stored in a packed array 
using the same time for 3 coloring a linked list so that processors can be used efficiently to the contracted lists.

In \cite{HanThesis,Han1} we showed that by resorting to the sublogarithmic 
CRCW array prefix sum algorithm \cite{Reif} we can contract linked lists in $O(n/p+\log n/\log^{(3)} n+\log l)$ time. In \cite{CV} it is improved to $O(n/p+\log n/\log \log n+\log l)$ time due to the fact that array prefix sum can be done in $O(n/p+\log n/\log \log n)$ time \cite{CV}. 

Previously we gave an $O(n\log i/p+\log^{(i)}n+\log i)$ time EREW PRAM 
algorithm \cite{Han} for finding a maximal independent set
or a maximal matching 
for linked lists or a 3-coloring of the linked lists, where $i$ can be any 
constructible integer. Here we use this algorithm and we apply some new ideas of ours and reach deterministic time $O(n\log i/p+(\log^{(i)} n +\log i)\log \log l+\log l)$ 
for linked lists contraction on the EREW PRAM, where $n$ is the total number of
nodes in all linked lists, $p$ is the number of processors used, $i$ is any 
constructible integer, $l$ is the length of the longest linked list.

In particular if $l=\Omega(\log^{(c)} n)$ for a constant $c$ we can 
reach $O(n/p+\log l)$ time. 

Our uniform linked list contraction algorithm will pack the linked list
of size $n/c$ for a constant $c$, after linked list 3 coloring and pairing-off nodes, into an
array of size $n/d$ for a constant $1 < d \leq c$, using the same time for
linked list 3 coloring. Thus, if linked list 3 coloring can be done in constant
time then the above linked lists contraction can be accomplished in $O(n/p+\log l)$ time for any values of $l$.

The result, that is uniform linked list contraction, has been envisioned before
but was thought probably unachievable. The result we have achieved here is
a surprise. We thought before that this is unachievable and are amazed of the
result achieved here. 
 
\section{Preliminary}

Linked list contraction is done by first finding a maximal independent
set of the linked list nodes and then let nodes in the maximal independent set
pair-off their neighboring node(s) with themselves (combine one's neighboring node and itself into one node and thus the name of contraction). An independent set
is a set of nodes such that there is no edge between any two nodes in the set.
A maximal independent set is an independent set such that it cannot be enlarged and remains to be an independent set. If two nodes $v_1$ and $v_2$ 
in the maximal 
independent set intend to pair-off the same (non)maximal independent set node $w$ then an arbitrary one of $v_1$ or $v_2$ wins and succeeds in the pairing-off with $w$. A maximal independent set can be found after linked lists nodes have been 3 colored. (Here 2 colored is OK also but 2 coloring is much slower than 3 coloring, use a maximum independent set is also OK but finding a maximum independent set is much slower than finding a maximal independent set). 

The problem in parallel computing for linked list contraction is where to 
place the contracted node. As in Anderson and Miller \cite{AM}, initially
we can imagine that the $n$ nodes linked list is placed in a two dimensional
array of $n/p$ rows and $p$ columns such that each of the $p$ processors
can be assigned to one column. Say processors initially working on the
first row of the array and proceeds downward the array. After the first
pair-off step in the worst case there are $n/2$ nodes in the first row after pairing-off because every pair of nodes before pairing-off are on the first
row. Thus we need to keep $p/2$ processors on the first row in the next step
and move the other $p/2$ processors to second row in the next step. Thus
in the worst case, nodes in some columns depletes faster than other columns.
This situation is shown in Fig. 1.
Before Anderson and Miller's work we do not even know that $n$ nodes will eventually be contracted to $n/\log n$ nodes as nodes in some columns may be all 
depleted and leave processors in these columns idling. Anderson and
Miller showed a dynamic load balancing strategy and a clever accounting scheme
to guarantee that the number of node be reduced to $n/\log n$ (actually
in their paper they showed that the number of nodes can be reduced to $n/\log^{(c)} n$ for a constant $c$, but this has the some effects as reducing to $n/\log n$ nodes and therefore here we talk about the situation of reducing to $n/\log n$
nodes). However, the remaining $n/\log n$ nodes are in the array with
some columns have more nodes than other columns. 
Anderson and Miller's algorithm cannot guarantee that the
$n/\log n$ nodes are placed in the remaining array of $n/(p\log n)$ rows and
$p$ columns without using extra $\log n$ time (resorting to packing). In Anderson and Miller's algorithm we need to spend $\log n$ time to pack the remaining
nodes to the array of $n/(p\log n)$ rows and $p$ columns. Packing takes
at least $\log n$ time on the EREW PRAM. On the CRCW PRAM it can be done
in $O(\log n/\log \log n)$ time \cite{CV}. Actually approximate packing 
(i.e. pack the remaining $n/\log n$ nodes to an array of size $cn/\log n$
for some small values of $c$, e.g. $c=\log\log n$, is ok also) and there
are algorithms for achieving approximate packing in $O(\log \log n)$ time
\cite{GZ} on the CRCW model.

\begin{figure}[htbp]
\epsfysize=2in 
\centerline{ \epsffile{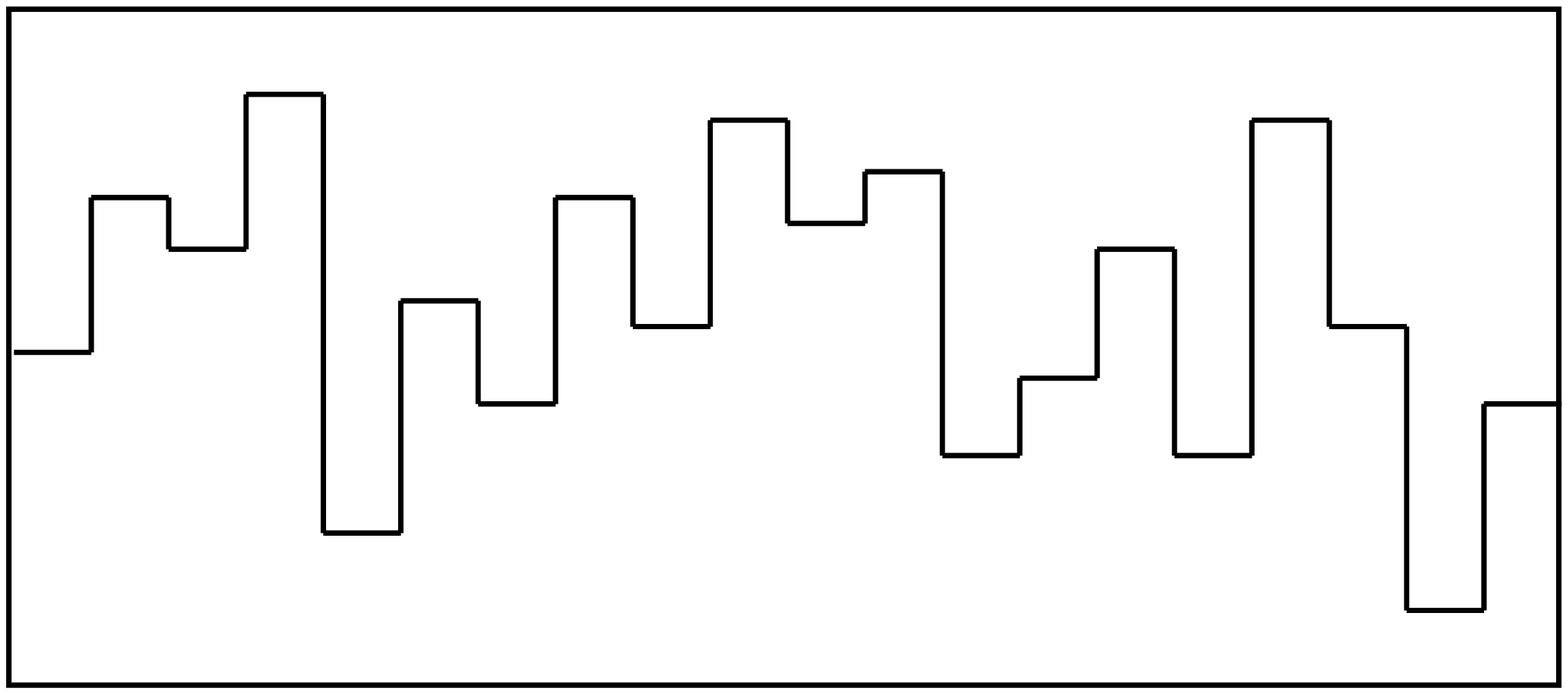}}
\centerline{Fig. 1. Remaining nodes after pairing-offs. }
\end{figure}

What we are actually demonstrate here is that after finding a maximal 
independent set a constant times and pairing-off a constant number of
passes (therefore there are $n/c$ contracted nodes) we can 
expending $O(n/p+1)$ time with $p$ processors 
to place these $n/c$ nodes into an array of
size $n/d$ with $c \geq d > 1$. 
This is what we call uniform linked list contraction. Thus for multiple linked
lists of total $n$ nodes and longest linked list has $l$ nodes. We can use
$n/\log l$ processors and contract
$n$ nodes to $n/\log l$ nodes and {\bf have these $n/\log l$ nodes
placed in an array of size $O(n/\log l)$ in $O(\log l)$ time }. This is
what has not been achieved before.    

In a sense to achieve uniform linked list contraction we just need to find
an orientation for the remaining paired-off nodes. As an example, say initially
we place $n$ nodes into an array of 2 rows and $n/2$ columns. After computing
a 3-coloring of the linked lists we can pair-off nodes colored with 2 with
one of its neighbors (one of its neighbor must be colored with 0 and the other
neighbor must be colored with color 1 for otherwise both of its neighbors
are colored with 0 and we can change its color to 1, or both of its neighbors
are colored with 1 and we can change its color to 0). Say we can move
nodes so that 
the two nodes in every column are colored with the same color (this can be easily achieved by sorting nodes in a constant number of consecutive columns by their colors), as shown in Fig. 2, if we get an orientation of the linked
lists, then pair-off the 0-1 pair along the orientation then we have one combined node remaining in each column. Previously such an orientation was thought
unachievable in constant time with $n$ processors.  
 
\begin{figure}[htbp]
\epsfysize=5in 
\centerline{ \epsffile{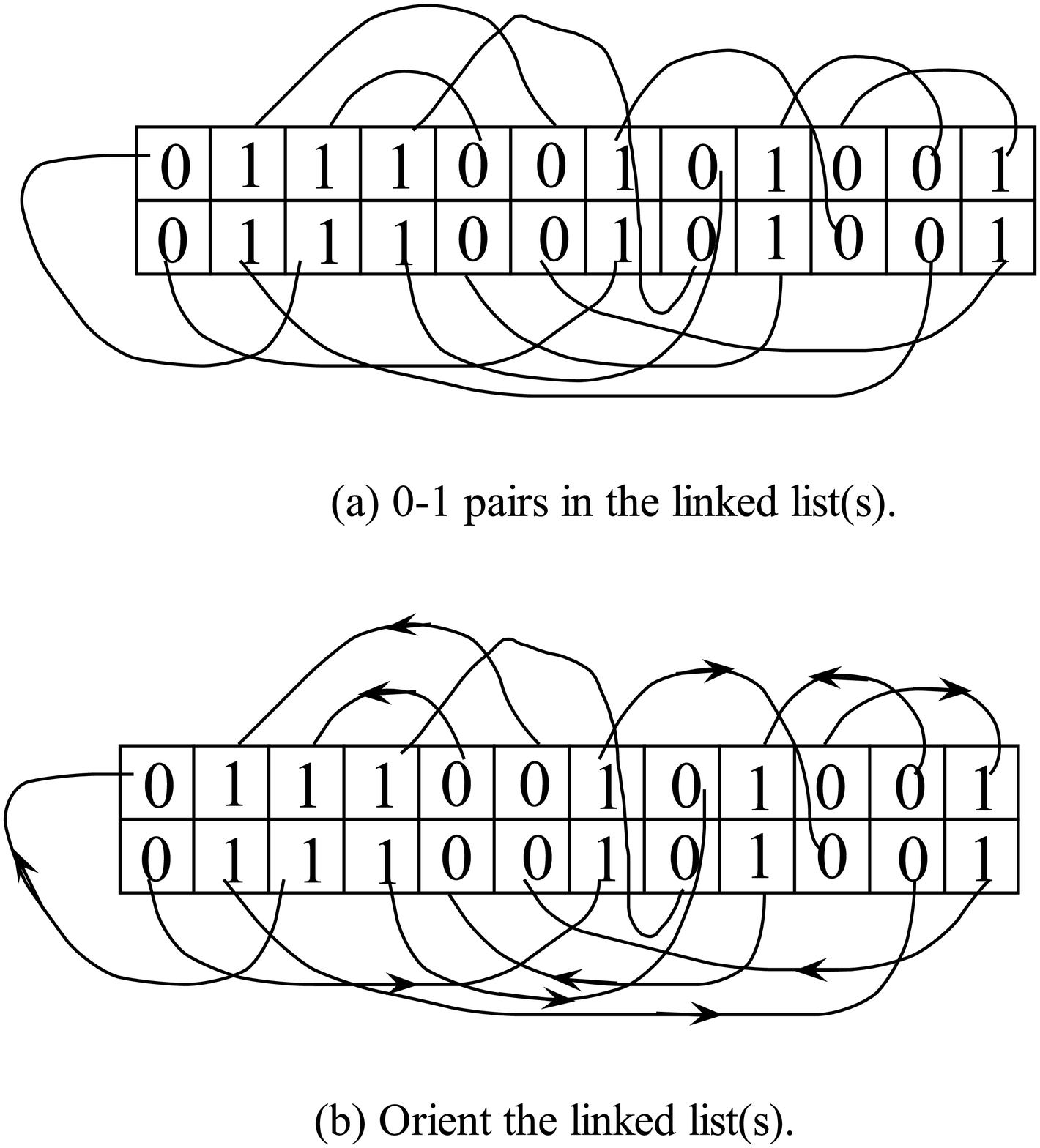}}
\centerline{Fig. 2. Uniform linked list contraction by orienting the linked list(s).}
\end{figure}

As linked lists are contracted they remain in this array $A$ although many cells in $A$ are empty now. In order to design an efficient parallel 
algorithm we need to pack the remaining linked lists to a smaller size array.
However, this packing operation is painful. We may assign 0's to the remaining cells containing contracted linked lists nodes and 1's to empty cells (where
the node of the linked list is paired-off with other node of the list) and then 
sort array elements by these 0's and 1's in order to pack the remaining linked
lists nodes to the front of the array. As is well known that this 0-1 sorting can be done by array prefix sum and it takes $O(n/p+\log n)$ time on the
EREW PRAM (or $O(n/p+\log n/\log \log n)$ time on the CRCW PRAM), where
$p$ is the number of processors used. The additive factor $\log n$ on the EREW
PRAM is not a pleasant factor and prevents the achievement of an optimal algorithm for linked list contraction on
the EREW model. 

Anderson and Miller \cite{AM} came up with a very clever way of overcoming this
packing difficulty. They used a dynamic load balancing scheme to achieve the
$O(n/p+\log n)$ time on the EREW PRAM for linked list contraction. However, because they do not pack linked list they cannot guarantee that the remaining linked list nodes after contraction are assigned evenly to all processors. This situation can be better explained by considering contracting multiple linked lists with the longest list has length $l$. These linked lists can be ``contracted'' by pointer jumping
\cite{Wyllie} in $\log l$ time with $n$ processors. If Anderson and Miller algorithm \cite{AM} is used to contract these linked lists it can be done in $O(n/p+\log n)$ time, but it cannot be done in $O(n/p+\log l)$ time. And this
$O(n/p+\log l)$ time is what we are aiming for in this paper. Basically, we show that
we can pack the remaining contracted nodes of the linked lists 
in the same time for computing
a 3 coloring, and this is what we call uniform linked list contraction. 

The strategy of our algorithm is to place the contracted nodes in the right
position. Note that we can use repeated pairing-off to purposely remove some
nodes and create vacant spots. Our strategy uses the colors created in
finding a maximal independent set to orient the linked list(s). 

Initially we may assume that linked lists are stored in a two dimensional array $A$ of 2 rows and $n/2$ columns. Our algorithm will first make every 0-1 pair
in the linked lists (nodes colored with 2 will be contracted into the linked
lists) on the same row. Secondly for every 0-1 on the bottom row (top row)
and in columns $c_0$ and $c_1$ we enforce that the two nodes in top row (bottom row)
and columns $c_0$ and $c_1$ be colored with the same color (i.e. both be colored with 0's or both be colored with 1's). This is shown in
Fig. 3. As shown in Fig. 3. Such configuration results in an orientation of
the linked lists because it creates the periodic pattern of $\begin{array}{cccc}0& 0 & 1& 1\\ 0 & 1 & 1 &0 \end{array}$ along the orientation.

\begin{figure}[htbp]
\epsfysize=5in 
\centerline{ \epsffile{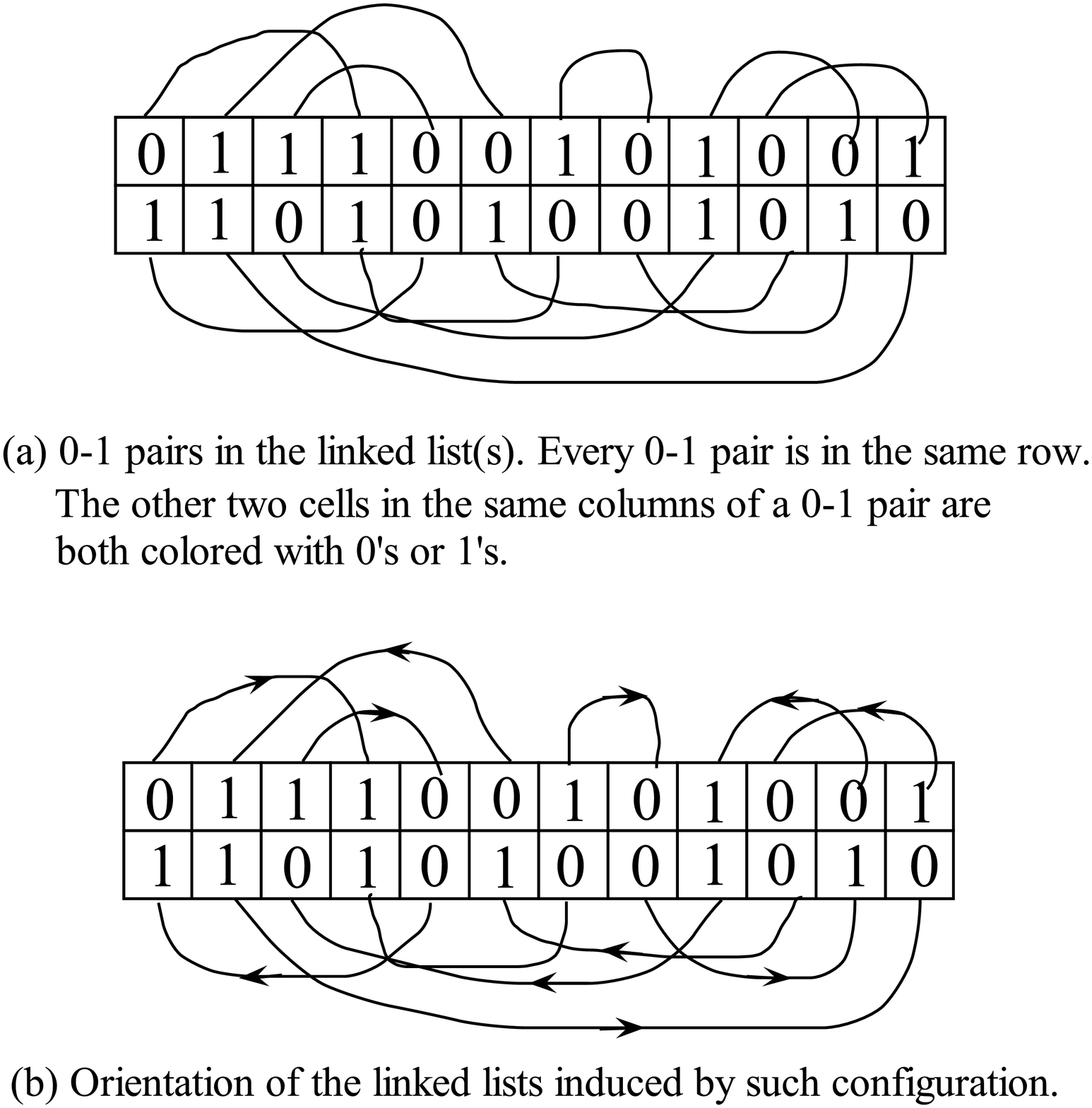}}
\centerline{Fig. 3. Create orientation for linked lists. }
\end{figure}

Thus, using the same time for 3 coloring a linked list, we can contract the linked list in $A$ and pack remaining nodes
of the linked list to the bottom row of $A$. 

The performance of our uniform linked list contraction algorithm is limited 
by the time for 3 coloring a linked list. If 3 coloring can be done in
$T$ time, then we can pack the contracted linked lists
into an array of size $n/d$ for a constant $d>1$ in $T$ time.

\section{The Algorithm}

\subsection{Localization}

We
use the localization technique. Localization 
basically says that we can separate linked lists into two or more (constant number of) parts such
that we can view that there are no links between any two of these parts (each
part is localized). Here we show how to store linked lists in two rows with each row containing $n/2$ nodes and then localize linked lists to each row (such that all links between two rows are deleted). In principle we can divide linked lists into any two parts and then localize them, i.e. we need not assume that any one part of the two parts are stored in a continuous block of memory. \\

\noindent
Algorithm {\bf Localize}

\begin{enumerate}
\item[1.] Store linked lists of $n$ nodes in a $2 \times n/2$ array $A$ such 
that each column has 2 nodes of the linked list. Call the 0th row 
as the upper level and 1st row as the lower level. We are going to localized the
upper row as well as the lower row.

\item[2.] Our first goal is to let a sufficient long (at least 100 nodes 
but no more than 300 nodes) of a sublist to run at the 
same level (either upper level or lower level). In order to achieve this 
we do (a): first contract sublist runs at the lower level and if a sublist 
has less than 100 nodes running at the lower level before it 
connects nodes at the upper level then we contract all these nodes 
and eliminate them by having them contracted to the nodes at the 
upper level. If the sublist running at lower level has at least 
100 nodes then we uncontract them and let this sublist run at the 
lower level. Then we do (b): contract sublist runs at upper level 
and if a sublist has less than 100 nodes running at the upper level 
before it connects nodes at the lower level then we contract all 
these nodes and eliminate them by having them contracted to the nodes 
at the lower level. If the sublist running at upper level has 
at least 100 nodes then we uncontract them and let this sublist 
run at the upper level but ``delete'' the links that 
connecting a node at lower level to an end node of this sublist
at upper level 
(there are 2 such links for each sublist). In this way we now 
have many (sub)lists running either at the upper level or 
at the lower level.
\end{enumerate}

The ``deletion'' of links connecting between upper level and lower level is
for the purpose of viewing that the linked lists are localized. It does not
mean that we will really cut any input linked list to multiple lists.
The localization is mainly to prevent linked list to alternatively going
up and down between the upper level and the lower level, i.e. prevent
sublist of the form $a_1, b_2, a_3, b_4, ...,$ where $a_i$'s are at the
upper level and $b_i$'s are at the lower level.

We can repeatedly apply this localizing algorithm to localize linked lists to a constant number of parts. For example, we can localize linked lists to 4 rows.

\subsection{Uniform Contraction}
We first localize the input linked list to two rows of $A$. Then we will 3 coloring the linked list on the bottom row. If a node is colored 2 instead of 0 or 1
then we will contract this node into one of its neighboring node. Thus now linked list on the bottom row is 0-1 colored. We then form 0-1 pairs. This is done
by letting each node colored with 1 arbitrarily pairs with one of its neighboring node colored with 0. If two nodes colored with 1 both try to pair with the same 0 colored node then we let the 1 node with larger address wins the pair. Nodes colored with 0 or 1 did not form pairs will be contracted into the neighboring nodes that formed pairs. Thus now we have the linked list at the bottom row of
$A$ having nodes paired into 0-1 pairs. 

We now do the same for the linked list at the top row of $A$ and form 0-1 pairs for nodes in the linked list on the top row of $A$. 

Now comes the key step. If two nodes at the bottom row $A[1, i]$ and $A[1,j]$
form a 0-1 pair we want to enforce that $A[0, i]$ and $A[0, j]$ at the top row
be colored with the same color (i.e. both be colored with 0's or both be 
colored with 1's). We call this the uniformity step. 
Note that, in general, $A[0, i]$ and $A[0, j]$
do not form 0-1 pair, for if $A[0, i]$ and $A[0, j]$ form a 0-1 pair then we
contract $A[0, i]$ and $A[0, j]$ into one node and place it at $A[1, i]$,
and we contract $A[1, i]$ and $A[1,j]$ into one node and place it at
$A[1, j]$.

To achieve the uniformity step, let us look at case here: $A[1, j_0]$ and $A[1, j_1]$ form a 0-1 pair and $A[0, j_0]$ and $A[0, j_1]$ are of different color. 
Say
$A[0, j_0]$ is colored with 0 and $A[0, j_1]$ is colored with 1. 
Now look at the 
pair of $A[0, j_1]$ and say this node is $A[0, j_2]$.
Let $A[1, j_2]$ and $A[1, j_3]$ form a 0-1 pair. $A[0, j_2]$ is colored with 0
as it is the pair of $A[0, j_1]$. Now look at the color of
$A[0, j_3]$. If $A[0, j_3]$ is colored with 0 then let $A[0, j_4]$ be the pair 
of $A[0, j_3]$ that is colored with 1. Let $A[1, j_4]$ and $A[1, j_5]$
form a 0-1 pair. If $A[0, j_5]$ is colored with 0 then we swap the position 
of $A[0, j_5]$ and $A[0, j_1]$ and thus accomplishing the 
uniformity step. If $A[0, j_5]$ is colored with 1 then we contract $A[0, j_1]$
and $A[0, j_2]$ into one node and place it at $A[0, j_2]$ and we move $A[0, j_3]$ to $A[0, j_1]$ and thus accomplish the uniformity step. 

If $A[0, j_3]$ is colored
with 1 then we form a chain (directed linked list) 
of $A[0, j_0], A[0, j_1], A[0, j_2], A[0, j_3], ...$,
where $A[0, j_{2i}]$'s are colored with 0 and $A[0, j_{2i+1}]$'s are colored with 1. $A[0, j_{2i+1}]$ and $A[0, j_{2i+2}]$ form a 0-1 pair. 
We group $A[0, j_{2i}]$ and $A[0, j_{2i+1}]$ into one node because
$A[1, j_{2i}]$ and $A[1, j_{2i+1}]$ is a pair. Thus, we get a directed linked list
$L$ of nodes $a_0, a_1, a_2, ...$, where $a_0$ represents $A[0, j_{0}]$ and $A[0, j_{1}]$, $a_1$ represents $A[0, j_{2}]$ and $A[0, j_{3}]$, .... We 3 color the nodes of $L$. If $A[0, j_{2i}]$ and $A[0, j_{2i+1}]$ is colored 0, 
  $A[0, j_{2i+2}]$ and $A[0, j_{2i+3}]$ is colored with 1,  $A[0, j_{2i+4}]$ and $A[0, j_{2i+1+5}]$ is colored with 2, then we swap $A[0, j_{2i+1}]$ and
$A[0, j_{2i+2}]$, contract $A[0, j_{2i+3}]$ and $A[0, j_{2i+4}]$ into one node
and place it at $A[0, j_{2i+4}]$, move $A[0, j_{2i+5}]$ to $A[0, j_{2i+3}]$. Thus accomplishing the uniformity step. If $A[0, j_{2i}]$ and $A[0, j_{2i+1}]$ is colored 0, 
  $A[0, j_{2i+2}]$ and $A[0, j_{2i+3}]$ is colored with 1,  $A[0, j_{2i+4}]$ and $A[0, j_{2i+1+5}]$ is colored with 0, then we swap $A[0, j_{2i+1}]$ and $A[0, j_{2i+2}]$ and thus accomplishing the uniformity step.

Note some pairs at the top row have been contracted into one node but this node
is the only node at the top row for the pairs at the same column and bottom row. That is, if $A[0, j_{2i+1}]$ and $A[0, j_{2i+2}]$ are contracted into one node
and placed at $A[0, j_{2k}]$ then $A[0, j_{2k+1}]$ is vacant. Thus, we can form 0-1 pairs for these contracted nodes (here $A[0, j_{2k}]$ and $A[0, j_{2k+1}]$) and need not worry about the uniformity
requirement.

Now we enforce the uniformity requirement for the bottom row. That is for each 0-1 pair at the top row $A[0, i]$ and $A[0, j]$ we require that $A[1, i]$ and
$A[1, j]$ be colored with the same color (both be colored with 0's or both be
colored with 1's). The approach is the same as we did
for the uniformity of the top row. 

After we enforced uniformity step, the colors of $2*color(A[0, j])+color(A[1, j])$ along the linked list form periodic pattern of $0, 1, 3, 2$ and thus it provides an orientation of the linked list $A[*, j], A[*, j+1], A[*, j+2], ...$, with links running
alternatively at the top and bottom rows. This linked list orientation then allow
us to contract the nodes at $A[0, j_{2i+1}]$ and $A[0, j_{2i+2}]$ into one node and place it at $A[1, j_{2i+2}]$ and contract nodes at $A[1, j_{2i}]$ and
$A[1, j_{2i+1}]$ into one node and place it at $A[1, j_{2i+1}]$.\\

\noindent
{\bf Main Theorem:} Linked lists can be uniformly contracted. $\framebox{}$\\

Because we can 3 color linked lists in $O(n\log i/p+\log^{(i)} n+\log i)$ time
and therefore if the longest linked list has length $l=\Omega (\log^{(c)} n)$ for 
a constant $c$ then we
can 3 color linked lists in $O(n/p+\log^{(c)}n)$ time. We can place input linked
lists on a two dimensional array of $\log^{(c+1)} n$ rows and $n/\log^{(c+1)}n$
columns. If we use $p=n/\log^{(c)} n$ processors then each row can be 3 colored
in $O(n/(p\log^{(c+1)} n))$ time and be contracted into the rows below. Thus, we eliminate
one row in $O(n/(p\log^{(c+1)} n))$ time. When the last row remains we can use pointer jumping to finish the linked lists contraction. Thus, it takes
$O(n/p+\log l)$ time to do the linked lists contraction.\\ 

\noindent
{\bf Corollary 1:} If multiple linked lists are placed into an array of size $n$ and the longest linked list has length $l=\Omega (\log^{(c)} n)$ for a constant $c$ then linked lists contraction
can be done in $O(n/p+\log l)$ time using $p$ processors on the EREW PRAM. 
$\framebox{}$\\

In general because we compute 3 coloring in $O(n\log i/p+\log^{(i)}n+\log i)$ time with $p$ processors, thus if we use $n/\log l$ processors then after $O(\log l \log i+(\log^{(i)}n+\log i)\log \log l)$ steps the size 
of the array containing all unfinished (contraction) linked lists has size $n/\log l$ and we can use pointer jumping to finish the linked lists
contraction in $O(\log l)$ time. Thus we have\\ 

\noindent
{\bf Corollary 2:} If multiple linked lists are placed into an array of size $n$ and the longest linked list has length $l$ then linked lists contraction
can be done in $O(n\log i/p+(\log^{(i)} n +\log i)\log\log l + \log l)$ time using $p$ processors on the EREW PRAM, where $i$ is a constructible integer. 
$\framebox{}$\\

\section{Conclusion}
We presented a deterministic algorithm for uniform linked list contraction. 
Because it is uniform and thus it brings some advantages such as for linked lists contraction.

\end{document}